\documentclass[12pt]{article}
  
\usepackage{amsmath}
\usepackage{amssymb}
\usepackage{bm}
\usepackage{cite}
\usepackage[dvips]{color}
\usepackage{graphicx}
\usepackage{slashed} \usepackage{tabularx}
\usepackage{hyperref}

\newcommand{\note}[1]{#1}

\begin{document}

\title{\Large\bf 
Probing the fourth generation Majorana neutrino dark matter 
}

\vspace{0.4truecm}
\author{
  Yu-Feng Zhou\footnote{Email: yfzhou@itp.ac.cn}\\ \\
            \textit{Kavli Institute for Theoretical Physics China,}\\
  \textit{State Key Laboratory of Theoretical Physics},\\ 
  \textit{Institute of  Theoretical Physics, Chinese Academy of Sciences}\\
  \textit{Beijing, 100190, P.R. China}
}
\date{}
\maketitle

\begin{abstract}
  Heavy fourth generation Majorana neutrino can be stable and contribute to a
  small fraction of the relic density of dark matter (DM) in the Universe. Due
  to its strong coupling to the standard model particles, it can be probed by
  the current direct and indirect DM detection experiments even it is a
  subdominant component of the whole halo DM.  Assuming that it contributes to
  the same fraction of the local halo DM density as that of the DM relic
  density in the Universe, we show that the current Xenon100 data constrain
  the mass of the stable Majorana neutrino to be greater than the mass of the
  top quark. In the mass range from 200 GeV to a few hundred GeV, the
  effective spin-independent cross section for the neutrino elastic scattering
  off nucleon is insensitive to the neutrino mass and mixing, and is predicted
  to be $\sim 1.5\times 10^{-44} \mbox{ cm}^2$, which can be reached by the
  direct DM detection experiments soon.  In the same mass region the predicted
  effective spin-dependent cross section for the heavy neutrino scattering off
  proton is in the range of $2\times 10^{-40} \mbox{ cm}^2\sim 2\times
  10^{-39}\mbox{ cm}^2$, which is within the reach of the ongoing \note{DM
    indirect search experiments such as IceCube.} We demonstrate such
  properties of the heavy neutrino DM  in a fourth generation model with the stability of the fourth
  Majorana neutrino protected by an additional generation-dependent $U(1)$
  gauge symmetry.
\end{abstract}

\newpage
\section{Introduction}

\indent Models with chiral fourth generation fermions are among the simplest and
well-motivated extensions of the standard model (SM) and have been extensively
studied~\cite{Frampton:1999xi}.  The condition for CP symmetry violation in
the SM requires at least three generations of
fermions~\cite{Kobayashi:1973fv}. However, there is no upper limit on the
number of generations from the first principle.
In the SM the amount of CP violation is not large enough to explain the
baryon-antibaryon asymmetry in the Universe. The inclusion of fourth
generation quarks leads to two extra CP phases in quark sector and possible
larger CP violation~\cite{0803.1234}, which is helpful for electroweak baryogenesis.  
With very massive quarks in the fourth generation, it has been proposed that
the electroweak symmetry breaking may become a dynamical feature of the
SM~\cite{Holdom:1986rn,Carpenter:1989ij,Hill:1990ge,Hung:2009hy}.

The recent LHC and Tevatron experiments have imposed constraints on the mass
of the fourth generation quarks from possible pair production processes.  For
instance, the lower limit on the mass of the fourth generation up-type quark
$u_4$ is found to be $m_{u_4} \geq 450\mbox{ GeV}$ from the search for the
process $u_4\bar{u_4} \to WbW\bar{b}\to b3j\ell^\pm \slashed{E}_T$, and the
limit on the mass of the fourth generation down-type quark $d_4$ is $m_{d_4}
\geq 490\mbox{ GeV}$ from the process $d_4\bar{d_4} \to WtW\bar{t}\to
\ell^\pm\ell^\pm b 3j \slashed{E}_T$~\cite{Contino}.
The direct searches for extra quarks at Tevatron have set lower limits on the
masses of  $u_4$ to be $m_{u4} \geq 335$ GeV, with the
assumption that the mass splitting between $u_{4}$ and the fourth generation down
quark $d_{4}$ is smaller than the $W$ mass and the branching ratio of
$u_{4} \to W q$ is 100$\%$~\cite{CDFnotePub}.  The lower limit for the mass of $d_{4}$
is $m_{d4} \geq 385$ GeV from the search for $d_{4}\to W t$ and the search for  $4W$ final states from the pair production of
$p\bar{p}\to d_4\bar{d}_4 $~\cite{Aaltonen:2011vr}, 
with the assumption
that $m_{u4}>m_{d4}$. Note that these limits are obtained
with the assumptions of maximum mixing between the fourth and the third 
generation quarks and $100\%$ branching ratio of the decay processes. 
The limits can be significantly weaker in models with suppressed mixing and decay branching ratio.

  In fourth generation models, the production of the Higgs boson can be
  enhanced by a factor of 5-9 due to the presence of two additional fourth
  generation quarks in the one-loop gluon-gluon fusion process.  In a combined
  analysis of ATLAS and CMS, the mass of the Higgs boson $h^0$ has been
  excluded in the range $120$-$600 \mbox{ GeV}$ which
  is based on the searches for $h^0\to \gamma\gamma$, $W^\pm W^\mp (W^{*\mp})$
  and $Z^0Z^0(Z^{*0})$ in fourth generation models ~\cite{Korytov}.  In obtaining
  such a bound, it is assumed that the fourth generation fermions are heavy
  and do not contribute to the total width of the Higgs boson. Note that for a
  Higgs boson heavier than 600 GeV, the self-interaction of the Higgs boson
  may be very strong and even nonperturbative,  while  a  Higgs boson lighter than 120 GeV may cause the
  problem of vacuum instability if the SM with fourth generation is valid up to
  the Planck scale.

The constraints on the masses of the fourth generation leptons are much
weaker.  The current lower bound on the mass of the unstable fourth generation
charged lepton $e_4$ is $m_{e4}\geq 100.8$ GeV from the search for $e_4$
decaying into the fourth generation neutrino $\nu_4$ and $W^\pm$ boson.  From
the invisible width of $Z^0$ boson, the lower bounds for the mass of an
unstable $\nu_4$ is set to be $m_{\nu4}\geq (101.3, 101.5,90.3)$ GeV from the
decay $\nu_4 \to (e,\mu,\tau) W^\pm$ in the case that $\nu_4$ is of Dirac
type. If $\nu_4$ is of Majorana type the corresponding bounds are modified to
be $m_{\nu4} \geq (89.5, 90.7, 80.5)$ GeV. The constraints are much weaker if
$\nu_4$ is a long-lived or stable particle. In this case the lower bound is
around $\sim m_Z/2$: $m_{\nu4}\geq 45.0(39.5)$ GeV for Dirac (Majorana)
neutrino~\cite{Nakamura:2010zzi}.  A long-lived fourth neutrino can also relax
the constraints from the precision electroweak
data~\cite{Murayama:2010xb}. If the fourth generation neutrino is stable, it can be a potential candidate
for the dark matter (DM) in the Universe.

Heavy stable neutrinos with mass greater than $\sim 1$ GeV are possible
candidates for the cold DM \cite{Lee:1977ua,Kolb:1985nn}.  However, if the
neutrino is the dominant component of the halo DM, the current DM direct
search experiments have imposed strong constraints on its mass.  For instance,
it has been shown that the spin-independent (SI) cross sections for the stable
Dirac neutrino elastic scattering off nucleus can be a few order of magnitudes
larger than the current DM direct search upper bounds due to the $Z^0$ and
$h^0$
exchanges~\cite{Goodman:1984dc,Srednicki:1986vj,Falk:1994es,0706.0526}. In the case of Majorana neutrino, the spin-dependent (SD) cross sections for
the elastic scattering can be very large.  Previous analysis based on the
assumption that the Majorana neutrino dominates the local halo DM density have
ruled out the mass range $10\mbox{ GeV}-2\mbox{ TeV}$ from the upper bounds on
the SD cross section ~\cite{Angle:2008we}. The mass of the Majorana neutrino is also constrained from SI scattering
through $h^0$-exchange if it has both Dirac and Majorana mass terms in the
flavor basis~\cite{Keung:2011zc}. 
On the other hand, it is well-known that for a neutrino heavier than $ \sim
m_Z/2$, the cross section for its annihilation is in general too large to
reproduce the observed DM relic density.  If the heavy neutrino mass is in the
range $ m_{Z}/2 \lesssim m_{\nu4} \lesssim m_W$, the annihilation into light
fermion pairs $f\bar{f}$ through $s$-channel $Z^0$ exchange contributes to a
very large cross section. For the neutrino heavier than $m_W$, the
contribution from $f\bar{f}$ channels decrease rapidly. However, other
channels such as $W^\pm W^\mp$, $Z^0 h^0$ etc.  are opened. For these
processes the corresponding cross section does not decrease with the
increasing of the neutrino mass, resulting in a relic density always decreases
with the growing of the neutrino mass, and a thermal relic density far below
the observed total DM relic density~\cite{Enqvist:1988we}.  
\note{ 
  Since the neutrino DM can only contribute to a small fraction of the
  relic density of DM, it is expected  that it  contributes to a small
  fraction of the halo DM density as well, and the two fractions are of the same order of magnitude. 
}
From model building point of view, it is easy to construct multi-component DM
models with the heavy neutrino being a subdominant component in the
halo. Despite its very low number density in the halo, it can still be probed
by the underground DM direct detection experiments due to its strong coupling
to the target nuclei, which provides a way to search for new physics beyond
the SM complementary to the LHC. In this case, the event rate of the
DM-nucleus elastic scattering will depend on both the relic density and
the cross section for the elastic scattering process. Since both of them have
nontrivial dependence on the  neutrino mass, the above mentioned
constraints on the neutrino DM could  be modified significantly.

In this work, we explore the consequence of this possibility in a model with a
fourth generation Majorana neutrino DM. The stability of the fourth Majorana
neutrino protected by an additional generation-dependent $U(1)$ gauge
symmetry. In the model the gauge-anomalies generated by the first three
generation fermions are canceled by the ones from the fourth generation.  We
perform an updated analysis of the Majorana neutrino DM in light of the recent
Higgs search results at the LHC and the recent DM direct detection results
such as Xenon100 and  SIMPLE  etc..  The relic density of the neutrino DM
is obtained by calculating the annihilation cross sections for all the
possible final states such as $f\bar{f}$, $W^\pm W^\mp$, $Z^0Z^0$, $Z^0h^0$ and 
$h^0h^0$ etc.. Under the assumption that the neutrino contributes to the same
fraction of the local halo DM density as that of DM relic density in the
Universe, we calculate the effective DM-nucleus elastic scattering cross sections
which are the cross sections rescaled by the fraction of the halo DM
density contributed by the heavy Majorana neutrino. 
The results  show that  the current Xenon100
data constrain the mass of the Majorana neutrino to be greater than $\sim
175\mbox{ GeV}$. In the mass range from $200\mbox{ GeV}$ to a few hundred
GeV, the spin-independent cross section for the neutrino elastic scattering
off nucleon is predicted to be $\sim 1.5\times 10^{-44}
\mbox{ cm}^2$, which is insensitive to the neutrino mass and can be reached by
the direct DM search experiments in the near future. In the same mass range
the predicted spin-dependent cross section for neutrino proton scattering is
in the range $2\times 10^{-40} \mbox{ cm}^2 \sim
2\times 10^{-39}\mbox{ cm}^2$, which is within the reach of the ongoing IceCube
experiment. Although the analysis 
is performed in a specific model, the results are valid for 
a wide range of models with heavy Majorana neutrino DM.

This paper is organized as follows, in section \ref{model}, we present the
details of the model with stable fourth generation Majorana neutrino and the
interactions relevant to the calculation of the relic density and DM-nucleus
elastic scattering cross sections. In section \ref{relic-density}, we
calculate the relic density of Majorana neutrino DM. In section
\ref{scattering}, we give the prediction for spin-independent and
spin-dependent cross sections for the heavy neutrino elastic scattering off
nucleon and compared them with the latest DM search experiments. Finally, we
give the conclusions in section \ref{conclusion}.

\section{
A model with stable fourth generation neutrino}\label{model}

In this work we consider a simple extension of the SM with a sequential fourth
generation and an additional $U(1)_F$ gauge symmetry. \note{The
  $U(1)$ extensions to the SM are well motivated from the point view of grand
  unification such as the $SO(10)$ and $E_6$ and have rich
  phenomenology~\cite{Langacker:2008yv} which can be reached by the on going  LHC
  experiments}. The flavor contents in the model are given by
\begin{align}   
q_{iL}=\begin{pmatrix} u_{iL} \\ d_{iL} \end{pmatrix}, \
\ell_{iL}=\begin{pmatrix} \nu_{iL} \\ e_{iL} \end{pmatrix}, \
u_{iR},\ d_{iR}, \ \nu_{iR}, \ e_{iR} \ (i=1,\dots, 4) .
\end{align}
All the fermions in the model are vector-like under the extra gauge
interactions associated with $U(1)_F$. The $U(1)_F$ charges of the fermions could  be  generation-dependent. In order to
evade the stringent constraints from the tree-level flavor changing neutral
currents (FCNCs), the $U(1)_F$ charges $Q_{qi}$ for the first three generation
quarks are set to be the same, i.e. $Q_{qi}=Q_q, \ (i=1,2,3)$ while $Q_{q4}=-3 Q_q$ for the
fourth generation quarks. Similarly, the $U(1)_F$ charges for the first three
generation and the fourth generation leptons are $Q_L$ and $-3 Q_L$,
respectively. In general, $Q_q$ and $Q_L$ can be different. For simplicity, in
this work we take $Q_q=Q_L=1$.
With this set of flavor contents and $U(1)_F$ charge assignments, it is
straight forward to see that the new gauge interactions are  anomaly-free. Since
the gauge interaction of $U(1)_F$ is vector-like, the triangle anomalies of
$[U(1)_F]^3$, $[SU(3)_C]^2 U(1)_F$ and $[\mbox{gravity}]^2 U(1)_F$ are all
vanishing.  The anomaly of $U(1)_Y [U(1)_F]^2$ is zero because the $U(1)_Y$
hypercharges cancel for quarks and leptons separately in each generation,
namely $\sum (- Y_{qL}+Y_{qR})=0$ and $\sum (- Y_{\ell L}+Y_{\ell R})=0$.  The
anomaly of $[SU(2)_L]^2 U(1)_F$ is also zero due to the relation
\begin{equation}\sum_{i=1}^{4}
Q_{qi}=0 \quad \mbox{and} \quad \sum_{i=1}^{4} Q_{Li}=0 .
\end{equation}
Thus in this model, the gauge anomalies generated by the first three
generation fermions are canceled by that of the fourth generation one, which
also gives a motivation for the inclusion  of the fourth generation.
 
The gauge symmetry $U(1)_F$ is to be spontaneously broken by the Higgs
mechanism. For this purpose we introduce two SM singlet scalar fields
$\phi_{a,b}$ which carry the $U(1)_F$ charges $Q_a=-2 Q_L$ and $Q_b=6 Q_L$
respectively. The $U(1)_F$ charges of $\phi_{a,b}$ are arranged such that
$\phi_a$ can have Majorana type of Yukawa couplings to the right-handed
neutrinos of the first three generations $\nu_{iR}\ (i=1,2,3)$ while $\phi_b$
only couples to the fourth generation neutrino $\nu_{4R}$. After the
spontaneous symmetry breaking, the two scalar fields obtain vacuum expectation
values (VEVs) $\langle \phi_{a,b}\rangle=v_{a,b}/\sqrt{2}$.

The relevant interactions in the model are given by
\begin{align}
\mathcal{L}&=
\bar{f}_i i\gamma^\mu D_\mu f_i 
+(D_\mu\phi_a)^\dagger (D_\mu\phi_a)+(D_\mu\phi_b)^\dagger (D_\mu\phi_b)
\nonumber\\
&-Y^d_{ij} \bar{q}_{iL} H d_{iR} -Y^u_{ij} \bar{q}_{iL} \tilde{H} u_{iR} 
-Y^e_{ij} \bar{\ell}_{iL} H e_{iR} -Y^\nu_{ij} \bar{\ell}_{iL} \tilde{H} \nu_{iR} 
\nonumber\\
&-\frac{1}{2}Y^m_{ij}\overline{\nu_{iR}^c} \phi_a \nu_{jR} \ (i,j=1,2,3)
-\frac{1}{2}Y^m_4 \overline{\nu_{4R}^c} \phi_b \nu_{4R} 
-V(\phi_a, \phi_b, H)+\mbox{H.c} .
\end{align}
where $f_i$ stand for left- and right-handed fermions, and  $H$ is the SM Higgs doublet.  $D_\mu f_i=(\partial_\mu -ig_1\tau^a
W^a_\mu -iYg_2 B_\mu -i Q_f g_F Z'_\mu)f_i$ is the covariant derivative with
$Z'_\mu$ the extra gauge boson associated with  the $U(1)_F$ gauge symmetry, 
and $g_F$ the corresponding gauge coupling constant. Since
$\phi_{a,b}$ are SM singlets, they do not play any role in the electroweak
symmetry breaking. Thus $Z'$ obtains mass only from the VEVs of the scalars
\begin{align}m_{Z'}^2=g_F^2 (Q_a^2 v_a^2+Q_b^2 v_b^2) ,
\end{align}
and it does not mix with $Z^0$ boson in mass term at tree level. It can only mix with $Z^0$
through  kinetic mixing. In this work we assume that the  effect of the kinetic mixing 
is small and negligible.
 
From the $U(1)_F$ charge assignments in the
model, the four by four Yukawa coupling matrix is constrained to be of the block diagonal
form $\bm 3\otimes \bm 1$ in the generation space. Since the $U(1)_F$ charges
are the same for the fermions in the first three generation, there is no 
tree level FCNC induced by the $Z'$-exchange in the
physical basis after  diagonalization.  Thus  a number of constraints from 
the low energy flavor physics such as the neutral meson mixings and the $b\to s \gamma$
can be avoided.

The contribution to the muon $g-2$ from the $Z'$ boson at one-loop can be estimated as~\cite{Baek:2001kca}
\begin{align}\Delta a_\mu\approx \frac{g_F^2}{12\pi^2}\frac{m_\mu^2}{m_{Z'}^2} .
\end{align} 
The current experimental data requires that $\Delta a_\mu \leq 3.9\times
10^{-9}$~\cite{Bennett:2006fi,Hagiwara:2006jt}, which can be translated into a
lower bound on the VEVs of $\phi_{a,b}$: $\sqrt{Q_a^2 v_a^2+Q_b^2 v_b^2}\geq
1.45\times 10^2 \mbox{ GeV}$. Note that the dependence on the coupling
constant $g_F$ is canceled by the one in the mass of $Z'$ in the expression of $\Delta
a_\mu$. 
The direct search for the process $e^+e^-\to Z'\to \ell^+\ell^-$ at
  the LEP-II leads to a lower bound on the ratio of the mass to the coupling to
  leptons: $M_{Z'}/g_{F} \geq 6 \mbox{ TeV}$~\cite{Carena:2004xs} for vector-like
  interactions, which corresponds to a more stringent lower bound:
  $\sqrt{Q_a^2 v_a^2+Q_b^2 v_b^2}\geq 6 \mbox{ TeV}$.

The current searches for narrow resonances in the Drell-Yan process
    $pp\to Z'\to \ell^+\ell^-$ at the LHC impose an alternative bound on the
    mass and the couplings of the $Z'$ boson.  In the narrow width
    approximation, the cross section for the Drell-Yan process can be
    parametrized as $\sigma =(\pi/(48s))[c_u w_u(s,M_{Z'}^2)+c_d
    w_d(s,M_{Z'}^2)]$, where $s$ is the squared center of mass energy and
    $w_{u,d}(s,M_{Z'}^2)$ are model-independent functions depending on
    $M_{Z'}$~\cite{Carena:2004xs}. The coefficients $c_{u,d}$ are related to
    the $Z'$ couplings to the quarks as $c_{q}=g^2(Q_{qL}^2+Q_{qR}^2)Br(Z'\to
    \ell^+\ell^-)\ (q=u,d)$ where $Q_{qL,qR}$ are the charges of the left- and
    right-handed quarks of a generic $U(1)$ gauge symmetry with coupling
    strength $g$.  The limit on the cross section reported by the experiments
    can be recast model-independently as limit contours in the $c_u-c_d$ plan
    with unique contour for each value of $M_{Z'}$~\cite{Carena:2004xs}.
    Using the $c_u-c_d$ contours, the limits on one type of $Z'$ model can be
    translated into that of other models.
    For a model with sequential neutral gauge boson $Z'_{SSM}$ which by
  definition has the same couplings as that for the SM $Z^0$
  boson~\cite{Langacker:2008yv}, the latest lower bounds on its mass
  $M_{Z'_{SSM}}$ is $1.94$ TeV from CMS~\cite{Timciuc:2011ji} and $1.83$ TeV
  from ATLAS~\cite{Collaboration:2011dca} respectively.
    In the $Z'_{SSM}$ model $c_u^{SSM}\approx 2.36\times 10^{-3}$ and $c_d^{SSM}\approx
  3.66\times 10^{-3}$.   The bound on $M_{Z'_{SSM}}$ can be translated into the bound on the mass
  and couplings of the $Z'$ in this model in which $c_u=c_d=18g_F^2 Br(Z'\to \mu^+\mu^-)$.
  By requiring that $c_u \leq c_{u}^{SSM}$ at $M_{Z'}= 1.94$ TeV, we obtain $g_F \leq 0.051$.
  Numerical analyses have shown that       for a given limit on the cross section, as the combination $c_u w_u +c_d
  w_d$ increases or decreases by an order of magnitude, the mass limit
  changes roughly by 500 GeV~\cite{Chatrchyan:2011wq,Accomando:2010fz}.  Thus
  as a rough estimation one can obtain limits on $g_F$ for other values of
  $M_{Z'}$, for instance $g_F\lesssim 0.029$ for $M_{Z'}=1.44$ TeV and $g_F
  \lesssim 0.0051$ for $M_{Z'}=0.94$ TeV, respectively.

\note{
  In this model, there is no mixing between the
  fourth and the first three generation quarks, thus the null results in searching
  for $u_4\to b W^\pm$ and $d_4\to t W^\pm$ at the LHC do not impose  constraints
  on the masses of $u_4$ and $d_4$.  The model is  not constrained by the 
  FCNC processes $u_4\to t X$ or $d_4\to b X$ either~\cite{Alwall:2011zm}.  
}

The fourth generation neutrinos obtain both Dirac and Majorana
mass terms through the vacuum expectation values (VEVs) of $H$ and $\phi_b$.  
In the basis of
$(\nu_L, \nu_R^c)^T$ the mass matrix for the fourth neutrino is given by
\begin{align}
m_\nu=\begin{pmatrix}
0 & m_D \\
m_D & m_M
\end{pmatrix} ,
\end{align}
where $m_D=Y_4^\nu v_H/\sqrt{2}$ with $v_H=246\mbox{ GeV}$ and $m_M=Y^m_4 v_{\phi_{b}} /\sqrt{2}$. 
The left-handed components  $(\nu_{1L}^{(m)}, \nu_{2L}^{(m)})$ of the two 
mass eigenstates are related to the ones in the  flavor eigenstates by a rotation angle $\theta$ 
\begin{align}
\nu^{(m)}_{1L}& =-i (c_{\theta} \nu_L -s_{\theta} \nu_R^c) , \quad
\nu^{(m)}_{2L} =s_{\theta}\nu_L +c_{\theta} \nu_R^c ,
\end{align}
where $s_{\theta}\equiv\sin\theta$ and  $c_{\theta} \equiv\cos\theta$. \note{The value of $\theta$ is defined in the range $(0,\pi/4)$} 
and  is determined by 
\begin{align}
\tan 2\theta=\frac{2m_D}{m_M},
\end{align}
with $\theta=0 \ (\pi/4)$ corresponding to the limit of minimal (maximal) mixing. 
The phase $i$ is introduced to render the two mass eigenvalues real and positive.
The two Majorana mass eigenstates are $\chi_1=\nu_{1L}^{(m)}+\nu_{1L}^{(m)c}$ and
$\chi_2=\nu_{2L}^{(m)}+\nu_{2L}^{(m)c}$, respectively.  The masses of the two neutrinos are given by
$m_{1,2}=(\sqrt{m_M^2+4 m_D^2}\mp m_M)/2$. In terms of the mixing angle $\theta$ they
can be rewritten as
\begin{align}
m_1=\left(\frac{s_{\theta}}{c_{\theta}}\right) m_D, \quad 
\mbox{and} \quad  m_2=\left(\frac{c_{\theta}}{s_{\theta}}\right) m_D ,
\end{align}
\note{ with $m_1\leq m_2$.  Note that for all the possible values of $\theta$
  the lighter neutrino mass eigenstate $\chi_1$ consists of more
  left-handed neutrino than the right-handed one, which means that $\chi_1$
  always has sizable coupling to the SM $Z^0$ boson. Therefore the LEP-II
  bound on the mass of stable neutrino is always valid for $\chi_1$, which
  is insensitive to  the mixing angle.}

\note{ As the fermions in the first three generations and  the fourth
  generation have different $U(1)_F$ charges, the fourth generation fermions
  cannot mix with the ones in the first three generations through Yukawa
  interactions.  After the spontaneous breaking down of $U(1)_F$, there exists
  a residual $Z_2$ symmetry for the fourth generation fermions which protect
  the fourth neutrino $\chi_1$ to be a stable particle if it is lighter than
  the fourth generation charged lepton $e_4$, which makes it a possible dark
  matter candidate.}

\note{ Although the stability of $\chi_1$ can be achieved easily by imposing an
  {\it ad hoc} $Z_2$ symmetry {\it only} on the fourth generation, it is
  theoretically more attractive to consider a $Z_2$ which originates from a
  $U(1)$ gauge symmetry in which all the fermions are charged.  First, models
  with extra $U(1)$ gauge symmetries are well motivated from various GUT
  models such as $SO(10)$ and $E_6$ etc., and it is common to link a discrete
  $Z_N$ symmetry to a broken continuous $U(1)$ symmetry in model building.
Second, in this model all the four generations are treated in parallel, The
only difference is the $U(1)$ charge of the fourth generation fermions. Due to
this $U(1)$ charge difference, after the spontaneous symmetry breaking by the
VEVs of $\phi_a$ and $\phi_b$ , the $U(1)$ is broken into two separate $Z_2$
symmetries, with one on the first three generation and the other one on the
fourth generation which stabilizes the fourth generation neutrino.
Third, in this model setup, all the fermions are vector-like under the extra
$U(1)$ gauge interactions. The cancellation of gauge anomalies automatically
requires the existence of the right-handed neutrinos in each generation. Thus
in this model it is natural to have  extra stable neutrinos.  }

In the mass basis the interaction between the massive neutrinos and the SM $Z^{0}$
boson is given by
\begin{align}
 \mathcal{L}_{NC}= \frac{g_1}{4\cos\theta_W}
  \left[
    -c_{\theta}^2 \bar{\chi}_1\gamma^\mu\gamma^5 \chi_1-s_{\theta}^2 \bar{\chi}_2\gamma^\mu\gamma^5 \chi_2
    +2i c_{\theta}s_{\theta} \bar{\chi}_1 \gamma^\mu \chi_2 
  \right]Z_\mu ,
\end{align}
where $g_1$ is the weak gauge coupling and $\theta_W$ is the Weinberg angle. 
The Yukawa interaction between $\chi_{1,2}$ and the SM Higgs boson is given by 
\begin{align}
\mathcal{L}_Y=-\frac{m_1}{v_H} \left(\frac{c_{\theta}}{s_{\theta}}\right)
\left[
c_{\theta}s_{\theta} \bar{\chi}_1\chi_1+c_{\theta}s_{\theta} \bar{\chi}_2\chi_2-i(c_{\theta}^2-s_{\theta}^2)\bar{\chi}_1 \gamma^5 \chi_2
\right] h^0 .
\end{align}
Since all the fermions are vector-like under the $U(1)_F$ gauge interaction and  $\chi_{1,2}$ are neutral particles, 
only the off-diagonal interaction $\bar{\chi}_1\chi_2Z'$ is allowed.

\section{Annihilation cross sections and relic density of the fourth 
generation  neutrino dark matter }\label{relic-density}

\begin{figure}[thb]
\begin{center}
\begin{tabular}{ccc}  
\includegraphics[width=0.25\textwidth,height=0.12\textwidth]{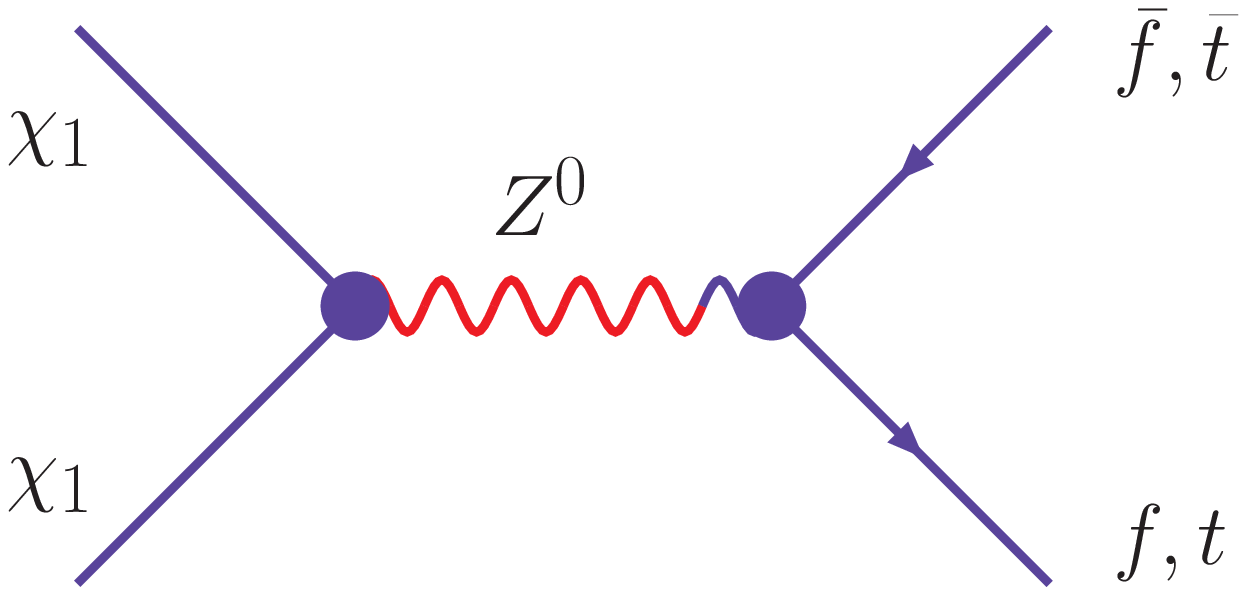}
&\includegraphics[width=0.25\textwidth,height=0.12\textwidth]{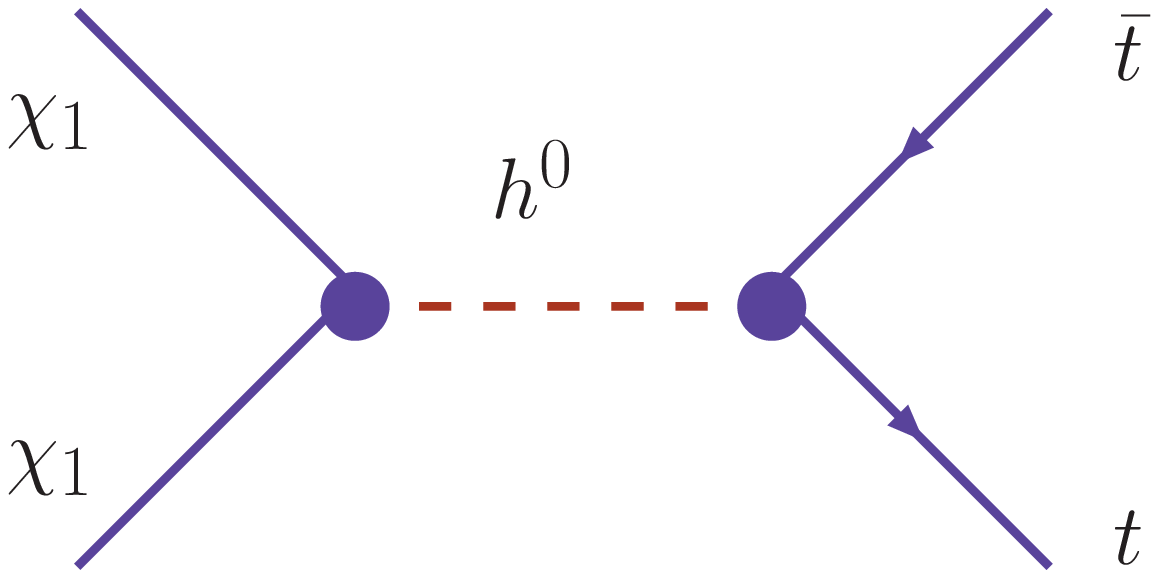} &\\
\includegraphics[width=0.25\textwidth,height=0.12\textwidth]{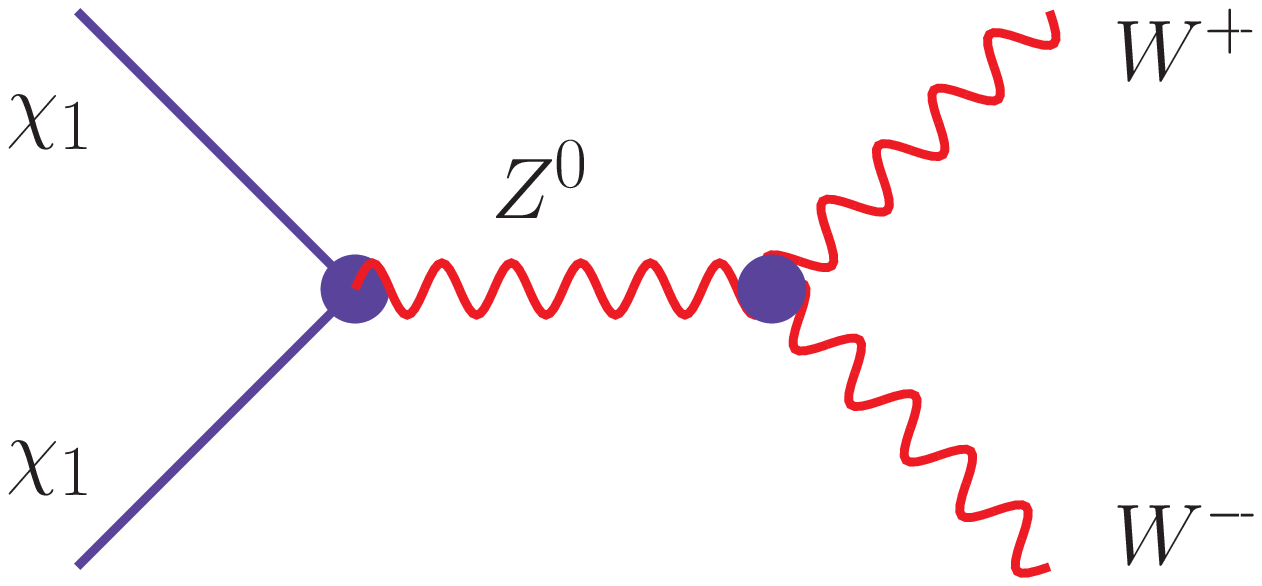}
&\includegraphics[width=0.25\textwidth,height=0.12\textwidth]{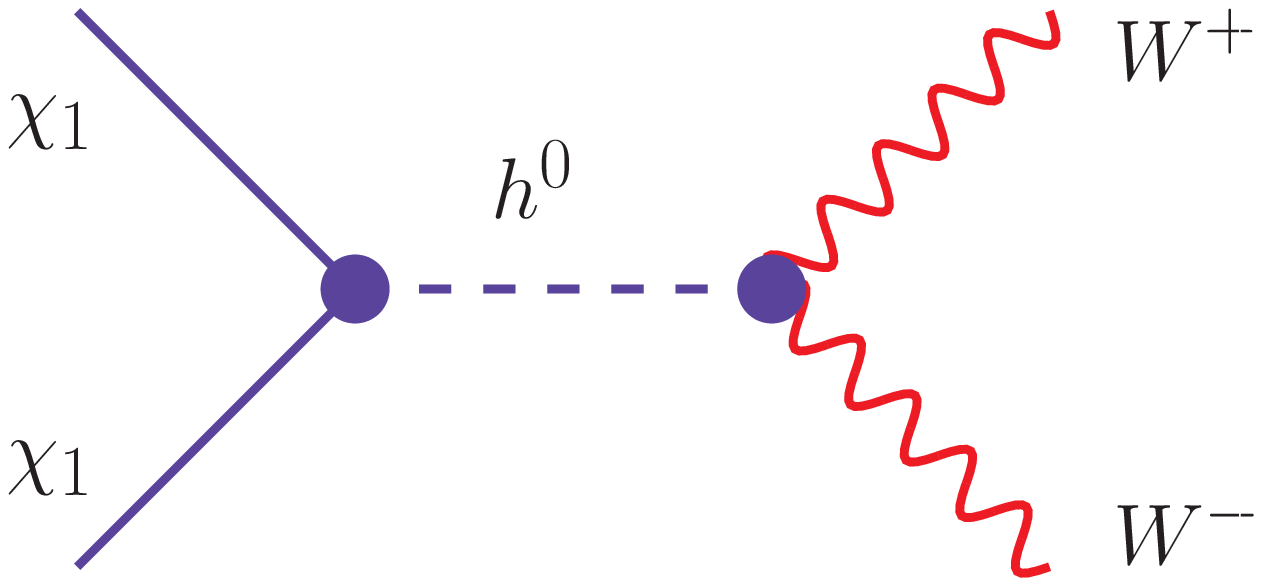}
&\includegraphics[width=0.25\textwidth,height=0.12\textwidth]{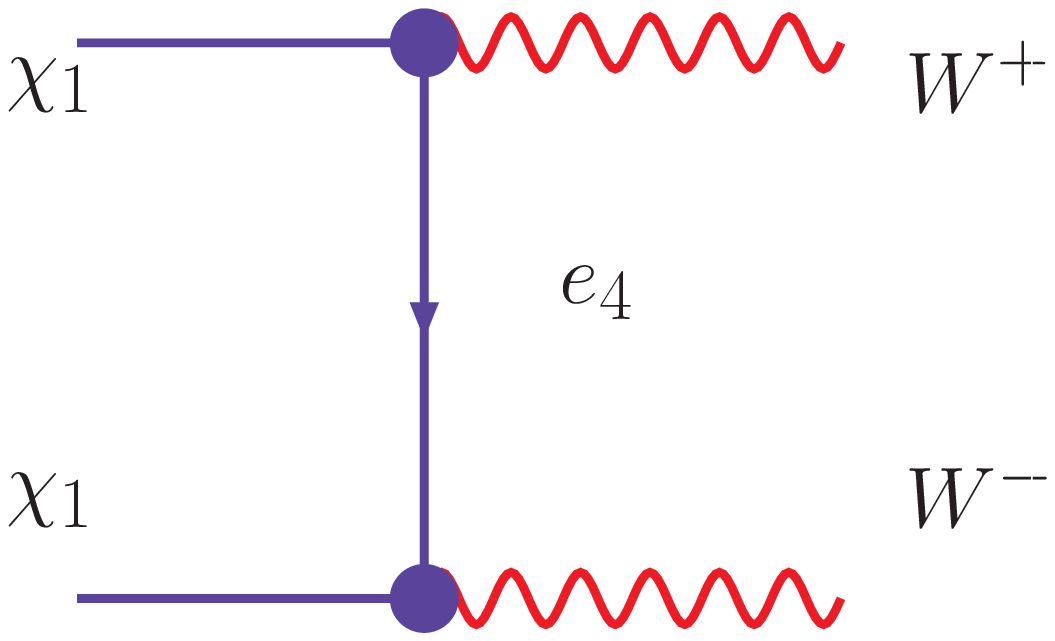}\\
\includegraphics[width=0.25\textwidth,height=0.12\textwidth]{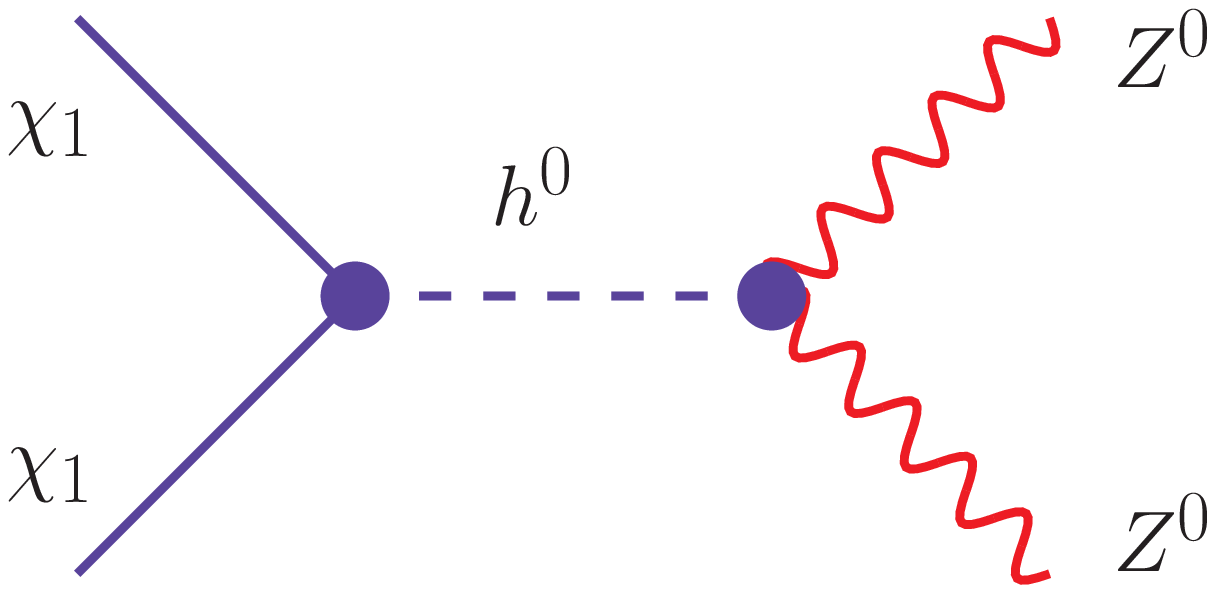}
&\includegraphics[width=0.25\textwidth,height=0.12\textwidth]{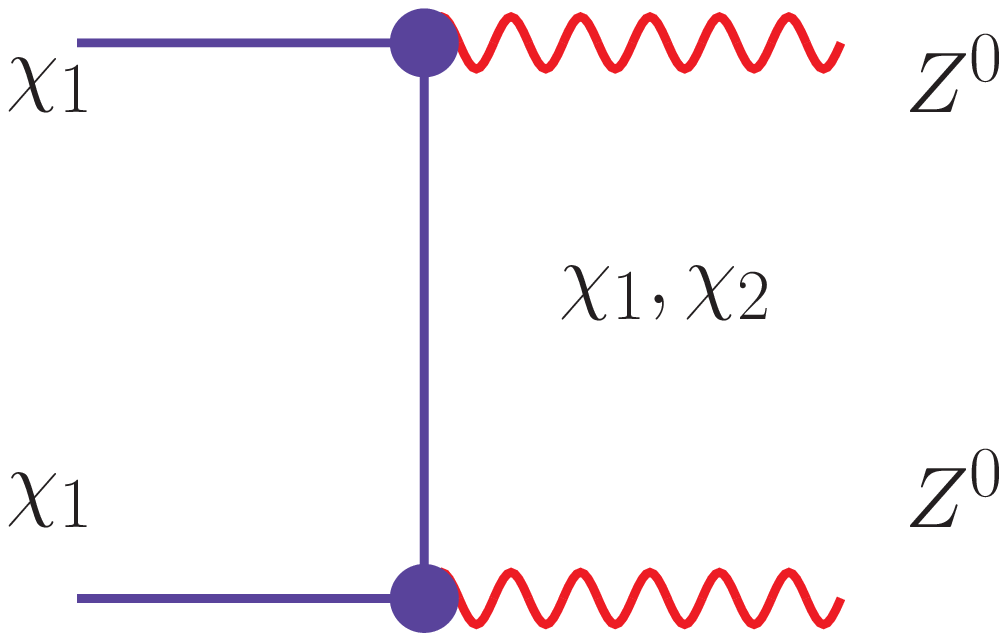}&\\
\includegraphics[width=0.25\textwidth,height=0.12\textwidth]{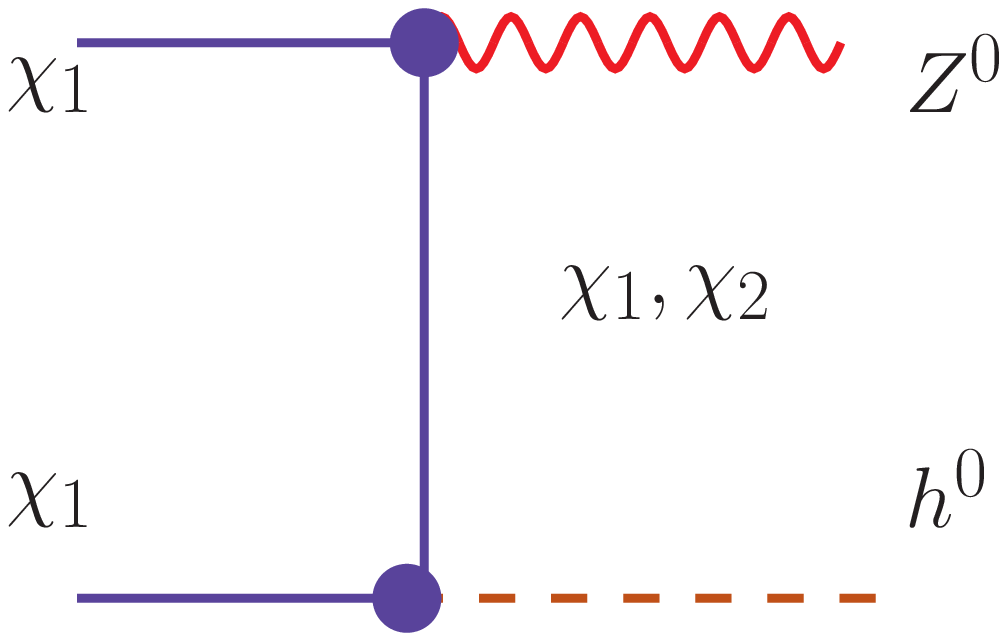}
&\includegraphics[width=0.25\textwidth,height=0.12\textwidth]{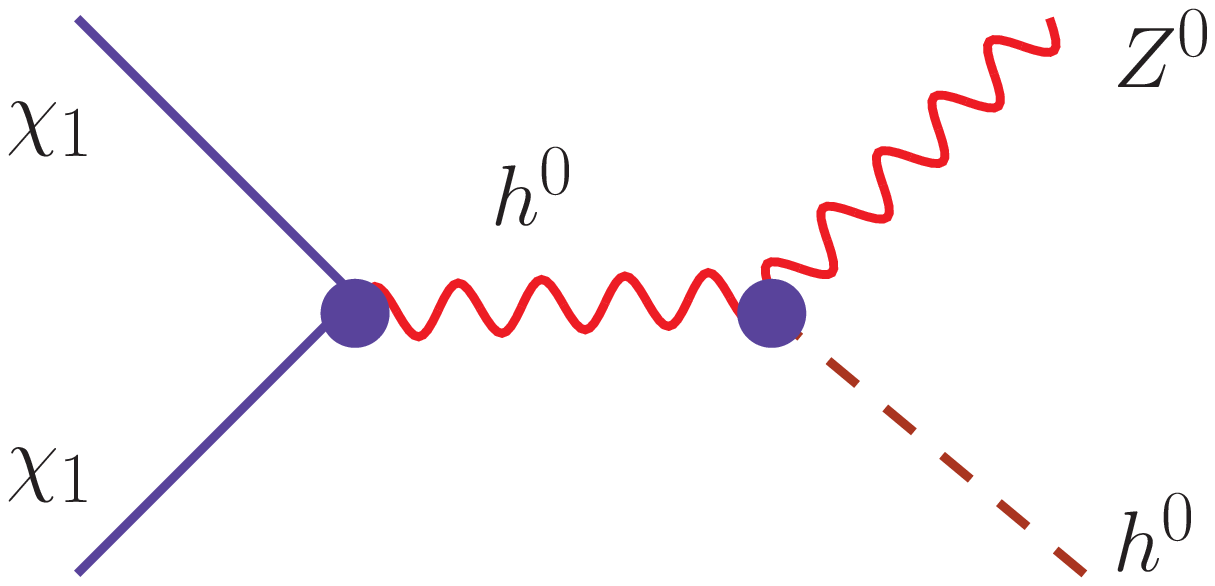}&\\
\includegraphics[width=0.25\textwidth,height=0.12\textwidth]{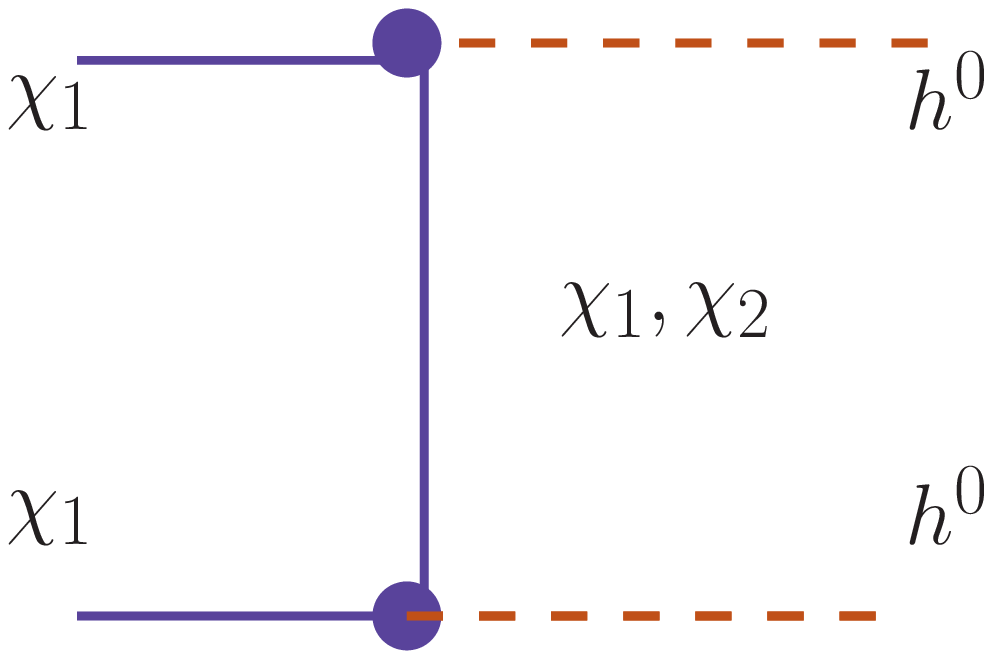}
&\includegraphics[width=0.25\textwidth,height=0.12\textwidth]{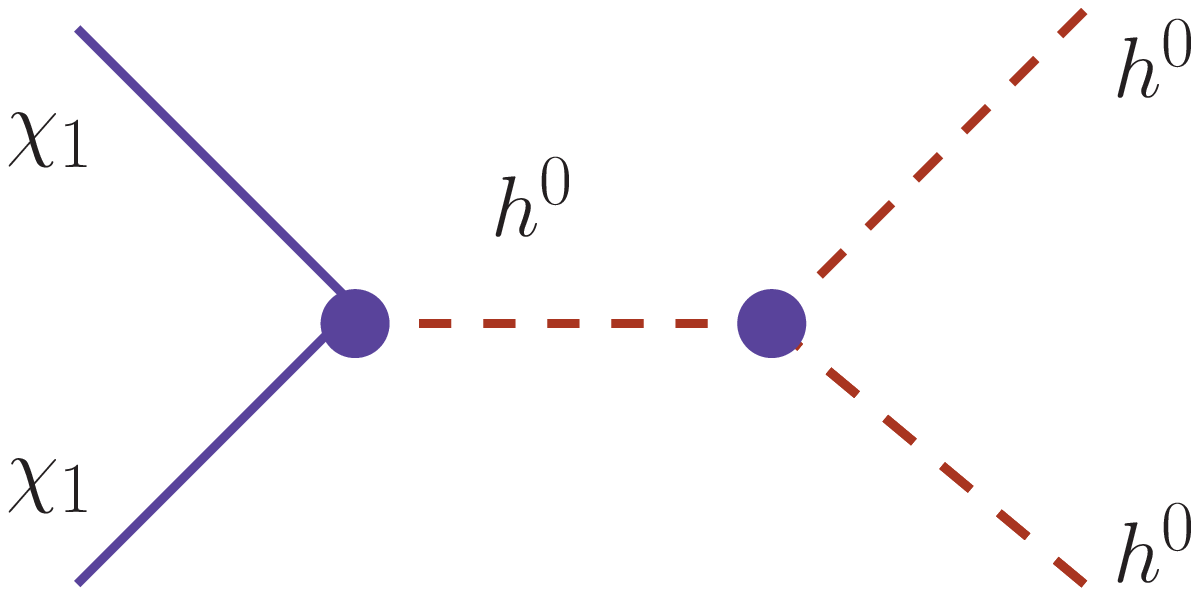}&
\end{tabular}
\caption{Feynman diagrams  for $\chi_1\chi_1$ annihilating into SM particles 
$\bar{f}f$, $\bar{t}t$, $W^\pm W^\mp$, $Z^0 Z^0$, $Z^0 h^0$ and $h^0 h^0$. }\label{fig:diagrams}
\end{center}
\end{figure}

The thermal relic density of $\chi_1$ is related to its annihilation cross section at
freeze out.  The Feynman diagrams for all possible
annihilation channels are shown in Fig. \ref{fig:diagrams}. When the mass of
$\chi_1$ is smaller than that of the $W^\pm$ boson, $\chi_1\chi_1$ can only
annihilate into light SM fermion pairs through $s$-channel $Z^0/h^0$ exchange.  For
Majorana neutrino the annihilation cross sections are suppressed by the small
masses of the final state fermions.  However, large enhancement of the
annihilation cross section occurs if the mass of $\chi_1$ is close to $m_Z/2$
such that the intermediate $Z^0$ is nearly on shell.
When the fourth generation neutrino is heavier than the $W^\pm$ boson, the
$W^\pm W^\mp$ channel will open and become important.  The 
process $\chi_1\chi_1\to W^\pm W^\mp$ involves  $s$-channel $Z^0$ and $h^0$ boson
exchange and $t$-channel $e_4$ exchange. All these three intermediate states
must be included in order  to maintain the  unitarity.
For even heavier $\chi_1$, the final states can be $Z^0Z^0$, $Z^0h^0$ and
$h^0h^0$.  The $Z^0Z^0$ channel contains the process of $s$-channel $h^0$
exchange and the $t$-channel $\chi_{1,2}$ exchange. Similarly, the
$Z^0h^0$ channel contains $s$-channel $Z^0$ exchange and the $t$-channel
$\chi_{1,2}$ exchange processes. The $h^0 h^0$ channel is dominated by the
$t$-channel processes due to the large Yukawa coupling.
When $\chi_1$ is heavier than the top quark, the $\bar{t}t$ final states must
be included. The cross section can be large as the top quark has  large
Yukawa couplings to the Higgs boson.
In this work we neglect the $Z'Z'$ final states by assuming that $Z'$ is always heavier than 
$\chi_1$.  

We numerically calculate the cross sections for $\chi_1\chi_1$ annihilation
into all the relevant final states using CalHEP 2.4 \cite{Belanger:2010gh}. In
order to determine the DM relic density, one needs to calculate the thermally
averaged product of the DM annihilation cross section and the relative velocity
\begin{align}
\langle \sigma v \rangle
=\frac{1}{8m_1^2 T K^2_2(m_1/T)}\int_{4m_1^2}^{\infty} ds \sigma (s-4m_1^2) \sqrt{s} K_1
\left(\frac{\sqrt{s}}{T} \right) , 
\end{align}
where  $T$ is temperature and $K_{1,2}(x)$ are the modified Bessel function of the second kind. 
The relic abundance can be approximated by
\begin{align}
\Omega h^2 \simeq \frac{1.07\times 10^9 \mbox{GeV}^{-1}}{\sqrt{g_*} M_{pl} \int_{x_F}^{\infty} \frac{\langle \sigma v \rangle}{x^2}dx} ,
\end{align}
where $x=m_1/T$ is the rescaled inverse temperature. $x_F\approx 25$ corresponds to the
decoupling temperature,  $g_*=86.25$ is the number of
effective relativistic degree of freedom at the time of  freeze out, and 
$M_{pl}=1.22\times 10^{19}$ GeV is the Planck mass scale.

\begin{figure}[htb]
\begin{center}
\includegraphics[width=0.45\textwidth]{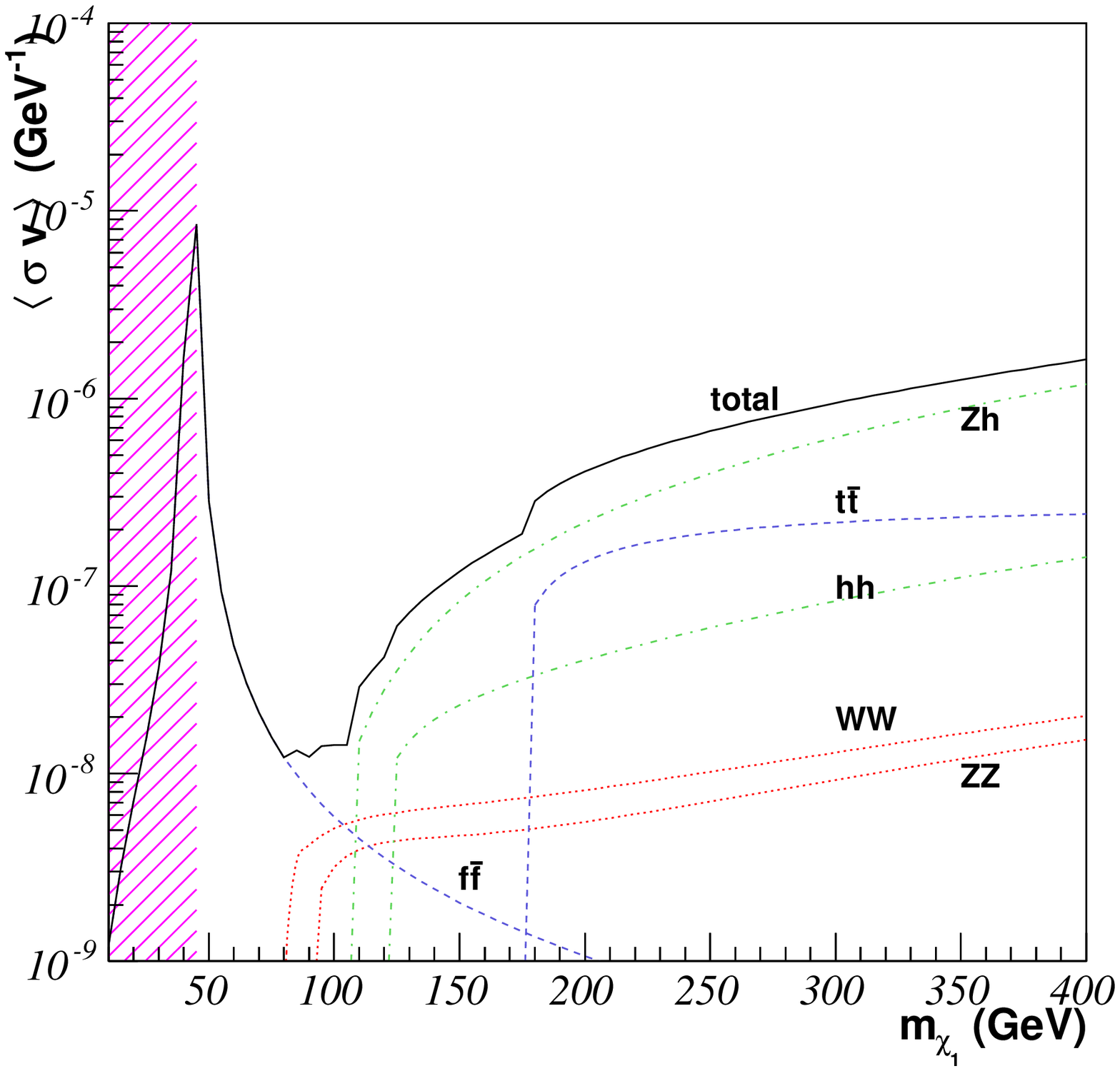}
\includegraphics[width=0.45\textwidth]{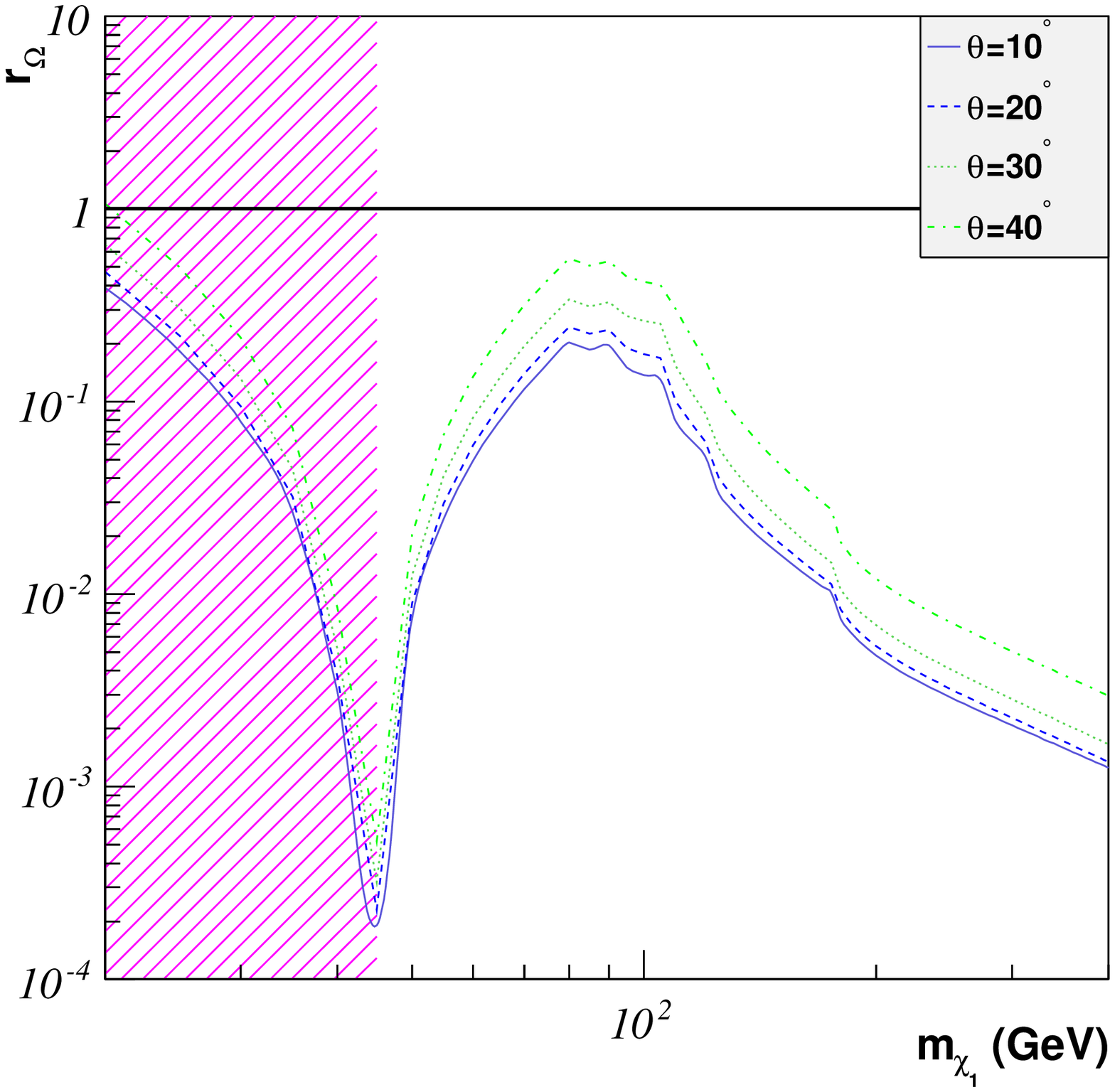}
\caption{Left) Thermally averaged product of the annihilation cross
  section and the relative velocity as function of the mass of the stable
  heavy neutrino $\chi_1$. The contributions from individual final states are
  also shown; right) the rescaled $\chi_1$ relic density  $r_\Omega$  as function of the mass of $\chi_1$.
 The shaded region is excluded by the LEP-II experiments.}
\label{fig:cross-section}
\end{center}
\end{figure}

In the left panel of Fig. \ref{fig:cross-section}, we show the value of
$\langle \sigma v\rangle$ at the time of freeze out $x_F=25$ as function of
$m_1$. In the figure, the contributions from individual final states are also
given. \note{The peak at $m_\chi\simeq m_{Z}/2$ corresponds to the case of
  resonant $s$-channel annihilation when the intermediate $Z^0$ boson is
  nearly on shell. } During the numerical calculations the mass difference
between the charged fourth generation lepton $e_4$ and $\chi_1$ is set to be
$m_{e_{4}}-m_{\chi_{1}}=50$ GeV. Note that a very large mass difference is
subject to strong constraints from the electroweak precision
data~\cite{Erler:2010sk} . The results are found to be insensitive to the
mass difference. In the left panel of Fig. \ref{fig:cross-section} the mixing
angle $\theta$ is fixed at $30^\circ$.  The recent LHC experiments
have placed strong constraints on the mass of $h^0$. For the fourth generation
model the mass of $h^0$ is constrained to be below 120 GeV or heavier than 600
GeV~\cite{Korytov}.  In the numerical calculations the mass of the SM Higgs
boson $h^0$ is fixed at 115 GeV.

In the case that the Higgs boson is light enough to be among the  final
states of the $\chi_1\chi_1$ annihilation, the contributions from the $Z^0h^0$
and $h^0h^0$ final states can be important. It has already been noticed that
the $Z^0h^0$ channel can be as importance as the $W^\pm W^\mp$ channel in the
case of heavy Dirac neutrino DM~\cite{0706.0526}. In the Majorana case, since the $W^\pm W^\mp$ cross section is strongly
suppressed by the smallness of the relative
velocity~\cite{Kainulainen:2006wq}, the $Z^0 h^0$ channel can give the
dominant contribution to the cross section.

In the right panel of Fig. \ref{fig:cross-section}, we show the quantity
\begin{align}  r_\Omega\equiv \frac{\Omega_{\chi_{1}}}{\Omega_{DM}}, 
\end{align}
which is the ratio of the relic density of $\chi_1$ to the observed total DM
relic density $\Omega_{DM} h^2=0.110\pm0.006$~\cite{Nakamura:2010zzi} as
function of the mass of $\chi_1$ for different values of the mixing angle
$\theta$. The results show a significant dependence on the mixing angle
$\theta$.  For smaller mixing angle $\theta$ the couplings between $\chi_1$
and gauge bosons $W^\pm, Z$ are stronger, resulting in a smaller relic
density. The results also clearly show that due to the large annihilation cross section,
$\chi_1$ cannot make up the whole DM in the Universe. $\chi_1$ can contribute
to $\sim 20-40\%$ of the total DM relic density when its mass is around 80
GeV.  But for $m_1\gtrsim m_t$, it can contribute only a few percent or less
to the whole DM.

However, since $\chi_1$ has strong couplings to $h^0$ and $Z^0$, even in the
case that the number density of $\chi_1$ is very low in the DM halo, it is
still possible that it can be detected by its elastic scattering off nucleus
in direct detection experiments.  Given the difficulties in detecting such a
neutral and stable particle at the LHC, there is a possibility  that the
stable fourth generation neutrino could be first seen at the DM direct
detection experiments.

\note{ In this model there exists three right-handed neutrinos
  $v_{iR}\ (i=1,2,3)$ in the first three generations. In the physical basis
  there may exist three sterile neutrinos, provided that the mixings between
  the left- and right-handed neutrinos are tiny.  If one of the sterile
  neutrinos has a mass around keV scale, it can be a good candidate for warm
  dark matter which could be  the dominant component of the DM in the
  Universe. The warm dark matter may provide a solution to some of the known
  problems in the DM simulations based on cold DM, such as reducing the number
  of subhalos and smoothing the cusps in the DM halo center.  In the SM with
  only right-handed neutrinos, the sterile neutrino may obtain the correct
  relic density through non-thermal production~\cite{Asaka:2005an}. In this
  model, since the keV sterile neutrino can annihilate into the light active
  neutrinos through the extra $U(1)_F$ gauge interactions, it can also be a
  thermal relic. In this case, the correct  relic density can be obtained by
  including the effect of entropy dilution through the decays of other heavier
  sterile neutrinos~\cite{Bezrukov:2009th}. }

\section{Direct detections of the fourth generation neutrino dark matter }\label{scattering}
The generic formula for the differential event rate of DM-nucleus scattering
per nucleus mass is given by
\begin{align}\label{eq:event-rate}
\frac{dN}{dE_R}=\frac{\rho_{DM} \sigma_N}{2m_{DM} \mu_N^2}F^2(E_{R})\int^{v_{esc}}_{v_{min}}d^3 v 
\frac{f(v)}{v} ,
\end{align}
where $E_R$ is the recoil energy, $\sigma_N$ is the scattering cross section
corresponding to the zero momentum transfer, $m_{DM}$ is the mass of the DM
particle, $\mu_N=m_{DM}m_{N}/(m_{DM}+m_N)$ is the DM-nucleus reduced mass,
$F(E_{R})$ is the form factor, and $f(v)$ is the velocity distribution function
of the halo DM. The local DM density $\rho_{DM}$ is often set to be equal to 
$\rho_0\simeq 0.3 \mbox{ GeV}/\mbox{cm}^3$ \note{(for updated determinations of $\rho_0$ see e.g. ~\cite{Catena:2009mf})} which is  the local DM density
inferred from astrophysics based on a smooth halo profile. 
\note{Since the neutrino DM can only contribute to a small fraction of the
  relic density of DM, it is likely that it also contributes to a small
  fraction of the halo DM density, namely, its local density $\rho_1$ is much smaller than $\rho_0$.  If the DM
  particles are nearly collisionless and there is no long range interactions
  which are different for different DM components, the structure formation
  process should not change the relative abundances of the  DM
  components.  In this work, we assume that $\rho_1$ is proportional to the
  relic density of $\chi_1$ in the Universe, namely}
\begin{align}r_\rho\equiv \frac{\rho_1}{\rho_{0}}\approx  \frac{\Omega_{\chi_{1}}}{\Omega_{DM}} ,
\end{align}
or $r_\rho\approx r_\Omega$. Consequently, the expected event rates of the DM-nucleus 
elastic scattering will be scaled down by  $r_\rho$.  In order to directly compare the theoretical
predictions with  the reported  experimental
upper limits which are often obtained under the assumption that the local DM particle density is
$\rho_0$, we shall calculate the rescaled elastic scattering cross section
\begin{align}\tilde{\sigma} \equiv   r_\rho \sigma \approx r_\Omega \sigma ,
\end{align}
which corresponds to the event rate to be seen at the direct detection experiments. 
Note that $\tilde{\sigma}$ depends on the mass of $\chi_1$ through the ratio $r_\rho$ even
when $\sigma$ is mass-independent. 

The spin-independent DM-nucleon elastic scattering
cross section in the limit of zero momentum transfer is given by~\cite{Jungman:1995df} 
\begin{align}
\sigma^{SI}_n=\frac{4 \mu_n^2}{\pi }\frac{[Z f_p+(A-Z) f_n]^2}{A^2} ,
\end{align}
where $Z$ and $A-Z$ are the number of protons and neutrons within the target
nucleus, respectively. $\mu_n=m_1 m_n/(m_1+m_n)$ is the DM-nucleon reduced mass. The coupling 
between DM and the proton (neutron) is given by
\begin{align}
f_{p(n)}&=\sum_{q=u,d,s}f^{p(n)}_{Tq} a_q \frac{m_{p(n)}}{m_q}
+\frac{2}{27} f^{p(n)}_{TG}\sum_{q=c,b,t}a_q \frac{m_{p(n)}}{m_q} ,
\end{align} 
with $f^{p(n)}_{Tq}$ the DM coupling to light quarks and $
f^{p(n)}_{TG}=1-\sum_{q=u,d,s}f^{p(n)}_{Tq}$. In the case that the elastic
scattering is dominated by $t$-channel Higgs boson exchange, the relation $f_n
\simeq f_p$ holds and one has $\sigma^{SI}_n \simeq 4 f_n^2\mu_n^2/\pi$. In
numerical calculations we take $f^p_{Tu}=0.020\pm0.004$,
$f^p_{Td}=0.026\pm0.005$, $f^p_{Ts}=0.118\pm0.062$, $f^n_{Tu}=0.014\pm0.003$,
$f^n_{Td}=0.036\pm0.008$ and $f^n_{Ts}=0.118\pm0.062$~\cite{hep-ph/0001005}.
The coefficient $a_q$ in the model is given by
\begin{align}
a_q=c_{\theta}^2\frac{m_1 m_q}{v_H^2 m_h^2} .
\end{align}
The value of $a_q$ is proportional to $m_1$, thus larger elastic scattering
cross section is expected for heavier $\chi_1$.  Note that in terms of $m_1$
the coefficient $a_q$ is proportional to $c_{\theta}^2$.  Part of the mixing
effects has been absorbed into the mass of $\chi_1$. In the limit of $\theta
\to 0$, $m_1$ is approaching zero and the couping between $\chi_1$ and $h^0$
is vanishing as expected. The value of $a_q$ has a strong dependence on
$m_h$. \note{As the latest LHC data exclude the mass
  of $h^0$ in the range $120 \mbox{ GeV}$-$600 \mbox{ GeV}$ in the
  presence of fourth generation fermions~\cite{Korytov}, we  fix
$m_h=115$ GeV in the numerical calculations.} The quark mass $m_q$ in the expression of
$a_q$ cancels the one in the expression of $f_{p(n)}$. Thus there is no quark mass dependence 
in the calculations.

\begin{figure}[htb]
\begin{center}
\includegraphics[width=0.65\textwidth]{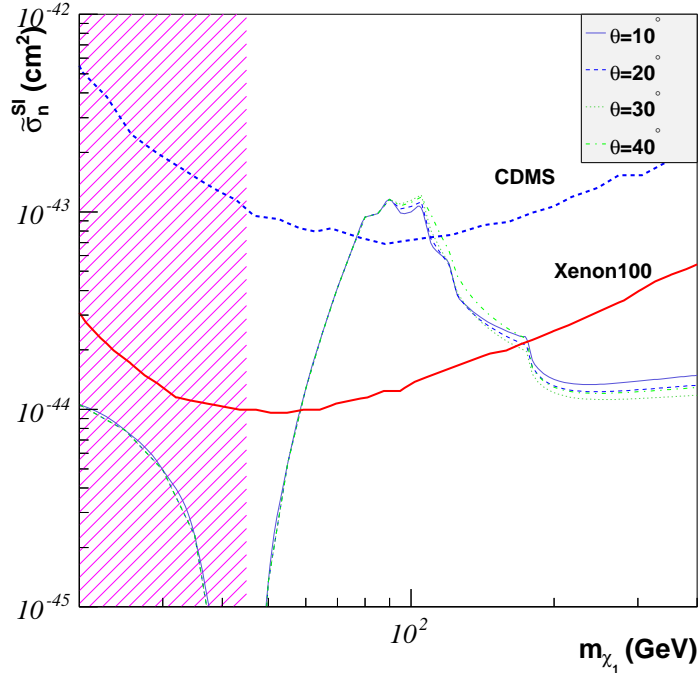}
\caption{Effective spin-independent cross section $\tilde{\sigma}^{SI}_n$
  which is $\sigma^{SI}_n$ rescaled by $r_\rho\approx r_\Omega$ for $\chi_1$
  elastically scattering off nucleon as function of the mass of $\chi_1$. Four
  curves correspond to the mixing angle $\theta=10^\circ$(solid),
  $20^\circ$(dashed), $30^\circ$(dotted) and $40^\circ$(dot-dashed)
  respectively. The current upper limits from CDMS~\cite{Ahmed:2009zw} and
  Xenon100~\cite{Aprile:2011hi} experiments are also shown.
}\label{fig:SIneutron}
\end{center}
\end{figure}

In Fig. \ref{fig:SIneutron} we give the predicted spin-independent effective cross
sections $\tilde{\sigma}^{SI}_n$ for the fourth generation neutrino 
elastic  scattering off nucleon as function of  its mass for
different values of the mixing angle $\theta$. One sees that even after the
inclusion of the rescaling  factor $r_\rho$, the current Xenon100 data can still
rule out a stable fourth generation neutrino  in the mass range 
$55\mbox{ GeV}\lesssim m_1 \lesssim 175 \mbox{ GeV}$ 
which corresponds to $r_\Omega \lesssim 1 \%$. Thus
the stable fourth generation neutrino must be heavier than the top quark,  and 
can only contribute to a small fraction of the total  DM relic density

On the other hand, for $m_{\chi_1} \gtrsim 200$ GeV, the cross
section does not decrease with $m_{\chi_1}$ increasing, and is nearly a
constant $\tilde{\sigma}^{SI}_{n}\approx 1.5\times 10^{-44}\mbox{cm}^2$ in the
range $200 \mbox{ GeV}\lesssim m_{\chi_1}\lesssim 400 \mbox{ GeV}$. This is
due to the enhanced Yukawa coupling between the fourth generation neutrino and
the Higgs boson which is proportional to $m_{\chi_1}$, as it is shown in the expression of 
$a_{q}$. 
One can see from the  Fig. \ref{fig:SIneutron}  that the result is not sensitive to the mixing
angle $\theta$ either, which is due to the compensation of the similar dependencies
on $\theta$ in the relic density. \note{For instance, the cross sections for the
$W^{\pm}W^{\mp}$ and $Z^{0}Z^{0}$ channel of  $\chi\chi$ annihilation are proportional to
$c_{\theta}^4$, which compensates the $\theta$-dependence in the $a_q$ for the
elastic scattering processes.}  

The Majorana neutrino DM can contribute to  spin-dependent elastic scattering  cross
section through  axial-vector interaction induced by the exchange of the
$Z^0$ boson. At zero momentum transfer, the spin-dependent cross section has
the following form~\cite{Jungman:1995df}
\begin{align}\sigma_N^{SD}=\frac{32}{\pi}G_F^2 \mu_n^2 \frac{J+1}{J}
\left(a_p \langle S_p\rangle + a_n \langle S_n\rangle \right)^2 ,
\end{align}
where $J$ is the spin of the nucleus, $a_{p(n)}$ is the DM effective coupling to 
proton (neutron) and $\langle S_{p(n)}\rangle$ the expectation value of the 
spin content of the nucleon within the nucleus. $G_F$ is the Fermi constant. 
The coupling $a_{p(n)}$ can be
written as
\begin{align}a_{p(n)}=\sum_{u,d,s}\frac{d_q}{\sqrt{2} G_F} \Delta^{p(n)}_q , 
\end{align}
where $d_q$ is the DM coupling to quark and $\Delta^{p(n)}_q$ is the fraction of the 
proton (neutron) spin carried by a given quark $q$. 
The spin-dependent DM-nucleon elastic scattering cross section is given by
\begin{align}
\sigma_{p(n)}^{SD}=\frac{24}{\pi} G_F^2 \mu_n^2 
\left(  \frac{d_u}{\sqrt{2} G_F }\Delta^{p(n)}_u
+\frac{d_d}{\sqrt{2} G_F }\Delta^{p(n)}_d
+\frac{d_s}{\sqrt{2} G_F }\Delta^{p(n)}_s
\right)^2  .
\end{align}
In numerical calculations we take 
$\Delta^p_u=0.77$, $\Delta^p_d=-0.40$, $\Delta^p_s=-0.12$\cite{Cohen:2010gj},
and use the  relations $\Delta^n_u=\Delta^p_d$, $\Delta^n_d=\Delta^p_u$,
$\Delta^n_s=\Delta^p_s$. The coefficients $d_q$ in this model are given by
\begin{align}d_u=-d_d=-d_s=\frac{G_F}{\sqrt{2}} .
\end{align}
For the axial-vector interactions, the coupling strengths do not 
depend on the electromagnetic charges of the quarks.

\begin{figure}[htb]
\begin{center}
\includegraphics[width=0.65\textwidth]{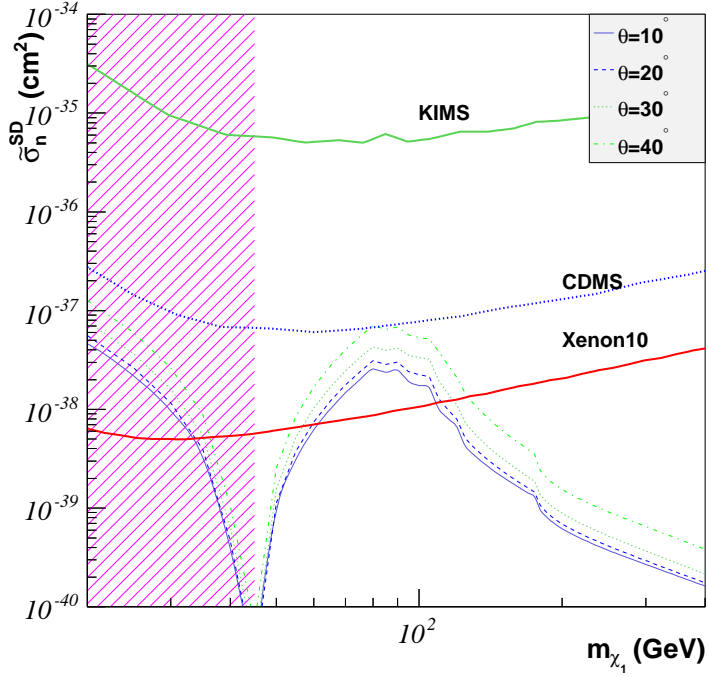}
\caption{ Effective spin-dependent cross section $\tilde{\sigma}^{SD}_n$ which
  is $\sigma^{SD}_n$ rescaled by $r_\rho\approx r_\Omega$ for $\chi_1$ elastically
  scattering off neutron as function of the mass of $\chi_1$. Four curves
  correspond to the mixing angle $\theta=10^\circ$(solid), $20^\circ$(dashed),
  $30^\circ$(dotted) and $40^\circ$(dot-dashed) respectively. The current
  upper limits from various experiments such as KIMS~\cite{Lee.:2007qn},
  CDMS~\cite{Akerib:2005za} and Xenon10~\cite{Angle:2008we} are also shown.  }
\label{fig:SDn}
\end{center}
\end{figure}

In Fig. \ref{fig:SDn} we show the predicted effective spin-dependent DM-neutron cross section
$\tilde{\sigma}_n^{SD}$ as function of the neutrino mass for different mixing
angles, together with various experimental upper limits.  Since
$\sigma^{SD}_n$ is independent of $m_{\chi_1}$, the dependency of
$\tilde{\sigma}_{p(n)}^{SD}$ on the neutrino mass comes from the dependency  of
$r_\rho$ on $m_{\chi_1}$, which can be seen by comparing Fig. \ref{fig:SDn}
with Fig. \ref{fig:cross-section}.
The Xenon10 data  is able to exclude the neutrino DM in the mass range
$60\mbox{ GeV}\lesssim m_{\chi_1} \lesssim 120\mbox{ GeV}$, which is not as
strong as that  from the Xenon100 data on spin-independent elastic
scattering cross section. For a heavy neutrino DM with mass in
the range $200\mbox{ GeV}\lesssim m_{\chi_1} \lesssim 400 \mbox{ GeV}$ the
predicted spin-dependent cross section is between $10^{-40}\mbox{ cm}^2$ and
$10^{-39}\mbox{ cm}^2$.

\begin{figure}[htb]
\begin{center}
\includegraphics[width=0.65\textwidth]{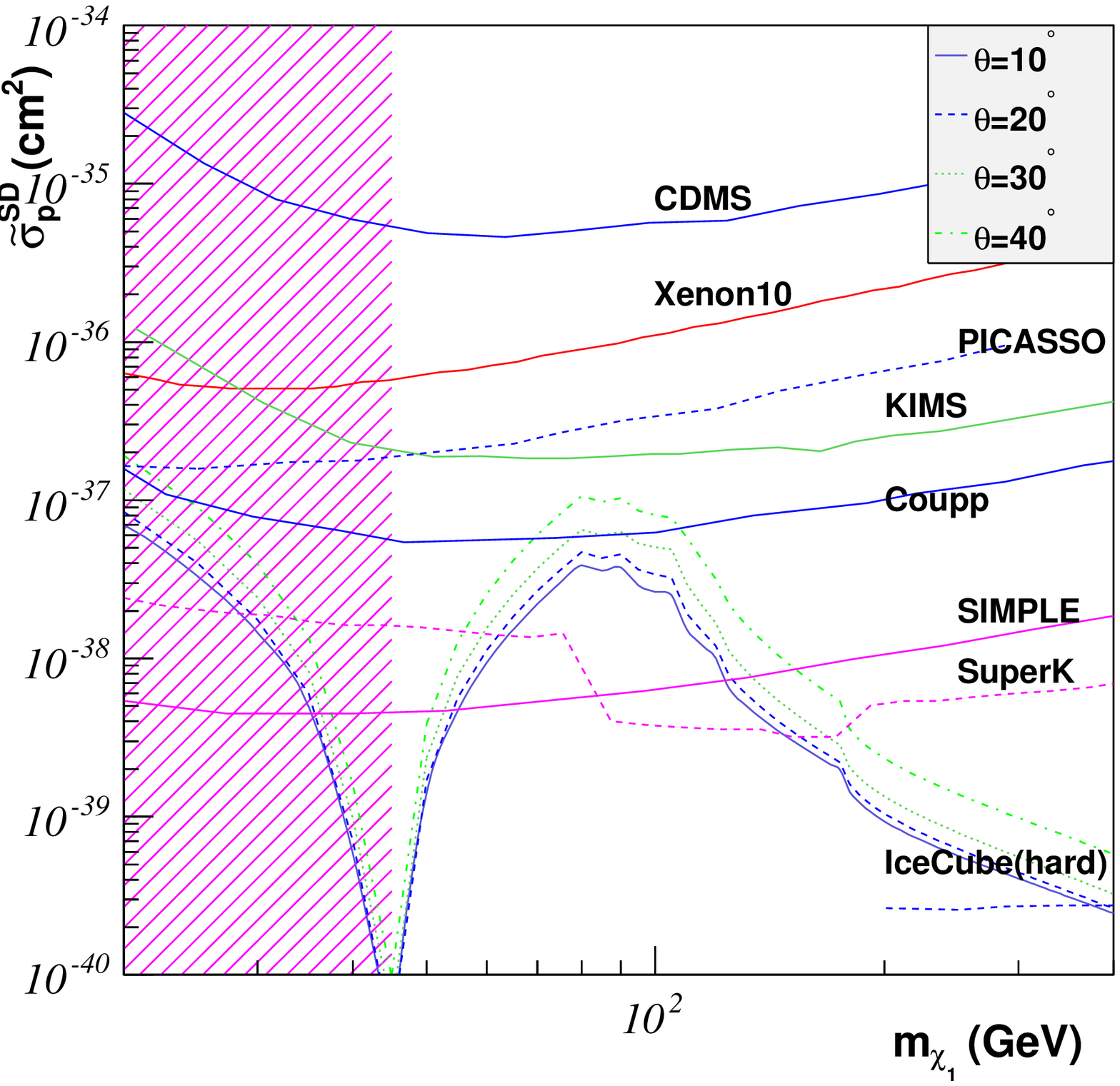}
\caption{Effective spin-dependent cross section $\tilde{\sigma}^{SD}_p$
which is $\sigma^{SD}_p$ rescaled by $r_\rho\approx r_\Omega$
  for $\chi_1$ elastically scattering off proton as function of the mass of
  $\chi_1$. Four curves correspond to the mixing angle
  $\theta=10^\circ$(solid), $20^\circ$(dashed), $30^\circ$(dotted) and
  $40^\circ$(dot-dashed) respectively.  The current upper limits from various
  experiments such as KIMS~\cite{Lee.:2007qn}, CDMS~\cite{Akerib:2005za},
  Xenon10~\cite{Angle:2008we}, Coupp~\cite{Behnke:2010xt},
  Picasso~\cite{Archambault:2009sm}, SIMPLE~\cite{Felizardo:2011uw},
  SuperK~\cite{Desai:2004pq}, and IceCube~\cite{0902.2460} are also shown.  }
\label{fig:SDp}
\end{center}
\end{figure}

In Fig. \ref{fig:SDp} we give the predicted spin-dependent DM-proton cross section
$\tilde{\sigma}_p^{SD}$. The cross sections for Majorana neutrino DM scattering off
proton and neutron are quite similar, which is due to the fact that the relative
opposite signs in $\Delta_u$ and $\Delta_d$ are compensated by the opposite
signs in $d_u$ and $d_n$. So far the most stringent limit on the DM-proton
spin-dependent cross section is reported by the SIMPLE
experiment~\cite{1106.3014}. The SIMPLE result is able to exclude the mass
range $50\mbox{ GeV}\lesssim m_{\chi_1} \lesssim 150\mbox{ GeV}$, which is compatible
with the constraints from Xenon100. 
In Fig. \ref{fig:SDp}, we also show the upper limits from indirect searches
using up-going muons which are related to the annihilation of stable fourth generation neutrinos captured in the
Sun. The limit from the Super-K experiment is obtained with the assumption
that 80$\%$ of the DM annihilation products are from $b\bar{b}$, $10\%$ from
$c\bar{c}$ and $10\%$ from $\tau\bar{\tau}$ respectively~\cite{Desai:2004pq}. In the range $170\mbox{ GeV} \lesssim m_{\chi_1} \lesssim 400\mbox{ GeV}$, the limit from
Super-K is $\sim 5\times 10^{-39}\mbox{ cm}^2$. The IceCube sets a stronger limit $\tilde{\sigma}_p^{SD} \leq 2\times 10^{-40}\mbox{ cm}^2$ for the DM mass at $250$ GeV~\cite{0902.2460}. 
This limit is obtained with the assumption that the DM annihilation products are 
dominated by $W^\pm W^\mp$. If the annihilation products are dominated by $b\bar{b}$, the 
limit is much weaker, for instance $\tilde{\sigma}_p^{SD} \leq 5\times 10^{-38}\mbox{ cm}^2$ for the DM mass at $500$ GeV~\cite{0902.2460}.  Note that in this model, the dominant final state is $Z^0h^0$. The expected limit should be 
somewhere in between. Nevertheless, the IceCube has the potential to test the predictions
in this model. 

\note{
  Different assumptions on the value of $r_\rho$  and the nature of the heavy stable neutrino may result in
  different limits.  For instance, in
  Ref.~\cite{Angle:2008we}, an excluded mass range of 10 GeV-2 TeV
  is obtained from the Xenon 10 data on the cross section of the
  spin-dependent DM-nucleus elastic scattering, which is based on the
  assumption that the local halo DM is entirely composed of stable Majorana
  neutrino, i.e. $r_\rho=1$, and the neutrino has the same couplings to the $Z^0$ boson as that
  of the SM active neutrinos. As in the present model we have $r_\rho \approx r_\Omega \ll 1$ and the coupling to the 
$Z^0$ boson depends on the mixing angle, the resulting constraints are different significantly.
 }

\section{Conclusions}\label{conclusion}
In conclusion, we have investigated the properties of stable fourth generation
Majorana neutrinos as dark matter particles.  Although they  contribute to a
small fraction of the whole DM in the Universe, they can still be easily probed by the
current direct detection experiments due to  their  strong couplings to the SM
particles.  We have considered a fourth generation model with the stability of
the fourth Majorana neutrino protected by an additional generation-dependent
$U(1)$ gauge symmetry. In the model the gauge-anomalies generated by the first
three generation fermions are canceled by the ones from the fourth generation.
We have shown that the current Xenon100 data constrain the mass of the stable
Majorana neutrino to be greater than the mass of the top quark. For a stable
Majorana neutrino heavier than the top quark, the effective spin-independent
cross section for the elastic scattering off nucleon is found to be
insensitive to the neutrino mass and is predicted to be around $10^{-44}
\mbox{ cm}^2$, which can be reached by the direct DM search experiments in the
near future.  The predicted effective spin-dependent cross section for the
heavy neutrino scattering off proton is in the range $10^{-40} \mbox{ cm}^2\sim
10^{-39}\mbox{ cm}^2$, which is within the reach of the ongoing \note{DM
  indirect search experiments such as IceCube.}

\section*{Acknowledgments}
The author is grateful to Yue-Liang Wu for encouragements and many helpful
discussions.  This work is supported in part by the National Basic Research
Program of China (973 Program) under Grants No. 2010CB833000; the National
Nature Science Foundation of China (NSFC) under Grants No. 10975170,
No. 10821504 and No. 10905084; and the Project of Knowledge Innovation Program
(PKIP) of the Chinese Academy of Science.

\providecommand{\href}[2]{#2}\begingroup\raggedright\endgroup
\end{document}